# GALAXY EVOLUTION FROM QSO ABSORPTION–SELECTED SAMPLES[*]

CHARLES C. STEIDEL[†][‡]
*MIT, Physics Department, Room 6-201*
*Cambridge, MA 02139, USA*

and

MARK DICKINSON
*Space Telescope Science Institute, 3700 San Martin Drive*
*Baltimore, MD 21218, USA*

## ABSTRACT

We present results of surveys for high redshift field galaxies selected by their having produced detectable absorption in the spectra of background QSOs. Such surveys, in essence selected by gas cross-section rather than by flux density, are almost completely independent of the conventional magnitude–limited redshift survey, and allow one to follow objects of normal luminosity well beyond $z \sim 1$, the redshift at which many of the standard techniques break down. We summarize the principal results for our completed survey at $z \leq 1$, and give the preliminary results of a second survey designed to extend the sample to $z = 1.6$. A general conclusion is that normal field galaxies exhibit strikingly little evolution in their space density, luminosity, and optical/IR colors to redshifts as high as $z \sim 1.5$. We discuss both the techniques involved and the implications of the results.

## 1. Introduction

It is slightly embarrassing, at a meeting dedicated in large part to the exciting current and planned instruments and observing programs designed to obtain very large numbers of faint galaxy redshifts, to be discussing a method of studying the evolution of galaxies that is probably *least* efficient in terms of the number of redshifts collected per unit observing time. On the other hand, the method of selecting galaxies through the features detected in a survey for QSO absorption lines certainly does satisfy the "wide field" aspect of the title of the meeting, since the galaxies which we will be discussed have been discovered along somewhat more than 150 different lines of sight, scattered all over the sky at appropriately high Galactic latitudes. In any case, as we attempt to demonstrate, the method is largely complementary to the types of projects discussed by others at this meeting, and it may represent our only hope of understanding both the broad evolutionary properties and the detailed physics of *normal* galaxies well beyond $z \sim 1$.

The limitations of optical redshift surveys for following normal galaxies beyond $z \sim 1$ are clear: if normal, relatively luminous galaxies were similar in the past, then a galaxy of characteristic $L^*$ luminosity ($M_B = -21.0$ for $H_0 = 50$ km s$^{-1}$ Mpc$^{-1}$) and mid–type spiral colors at $z = 1.2$ has $B = (24.7, 25.2)$ and $I = (22.7, 23.2)$ for $q_0 = (0.5, 0.05)$. In addition, the spectral features used as redshift determinants for faint galaxies (primarily [OII]$\lambda$3727 in emission and Ca II K and H/4000Å break in absorption for late–type and early–type galaxies, respectively) occur between 8200 and 8900 Å at the same redshift, a region where sky subtraction is increasingly problematic and detector sensitivity is falling off (see, however, the contribution by David Crampton in these same proceedings). Clearly even the simple *identification* of such a galaxy would be problematic, and one would pick up only the very brightest of the galaxy population. Ideally, one would not only like to *count* galaxies at high redshift, but also to obtain additional information which might provide clues to subtler forms of evolution, and to be able to follow in a detailed, physical sense, the nature of a "normal" galaxy back to the epoch of its formation. Therein lies the "niche" for a sample selected by QSO absorption line systems.

As we have discussed elsewhere[22,23,26,27], there are a number of advantages to a sample of galaxies selected by QSO absorption, as compared to one selected by apparent magnitude in a particular observed passband. One is that the galaxies are selected in essence by gas cross-section, as "flagged" by the presence of a detectable doublet of Mg II $\lambda\lambda$2796, 2803. A few general comments on this point: first, this spectroscopic feature can be observed from the ground (using identical *rest–frame* criteria) over the redshift range $0.2 \leq z \leq 2.2$, thereby connecting epochs about which we know practically nothing with the relatively "local" universe; second, this feature turns out to be an almost exact tracer of H I gas with column density $N(H\ I) \gtrsim 10^{17}$ cm$^{-2}$, so that the metallicity of the gas is not important–the presence of Mg II at the strength observed in the absorption line surveys indicates a certain quantity of H I, and that same quantity of H I is not found without accompanying detectable Mg II. What will turn out to be important in subsequent analyses below is that the redshift distribution of the Mg II systems is consistent with no significant change in the *total* cross-section (i.e., $n\sigma$, where $n$ is the object co-moving space density and $\sigma$ is the average cross-section of a single absorber) over the entire range observed (see Figure 1).

Because the sample is gas cross-section selected, there is (in principle) no obvious bias with respect to galaxy color, surface brightness, or luminosity and the level of incompleteness is always known, since the presence of an object at a particular redshift is known *a priori*, and one can simply search until it is found, or if it is not found, set limits on its properties. This means that one can probe a large range of luminosity, independent of redshift. Unlike the pencil–beam redshift surveys, the absorption–selected sample is relatively free of large–scale structure biases, since the galaxies are chosen over widely–separated lines of sight. Finally, what is possibly the most fundamental advantage to an absorption–selected sample in the long run, is that each galaxy has a bright QSO "behind" it with a range of impact parameters throughout the sample. While the presence of this object makes some aspects of the galaxy identification more difficult, it provides direct access, through high resolution spectroscopy of the QSOs, to detailed physical information on the galaxies, (e.g., chemical content, gas–phase physical conditions, gas–phase extent, structure, kinematics) with the richness of detail being essentially *independent of redshift*. For the present, however, we will concentrate on first establishing the population of galaxies that is selected using gas cross-section, comparing the results to the field galaxy redshift surveys over the redshift range common to both, and then using the statistical baseline at moderate redshift to investigate the evolutionary properties of "normal" galaxies to redshifts that are largely unexplored by the field redshift surveys.

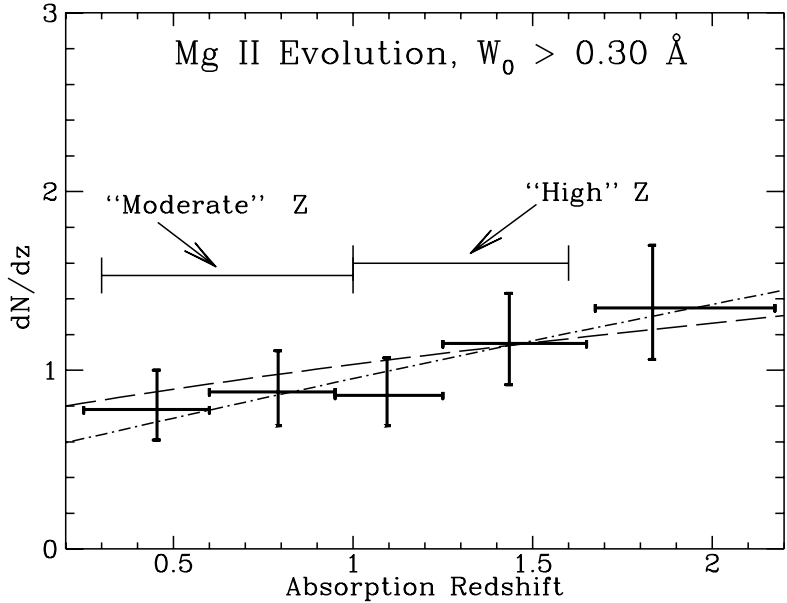

Fig. 1. Plot showing the number of Mg II absorption systems per unit redshift range versus redshift[28]. The expected behavior of the $dN/dz$ relation for no evolution in the total co–moving cross-section of the absorbers is shown with the dash–dot ($q_0 = 0.05$) and long–dashed ($q_0 = 0.5$) curves, respectively. The two redshift ranges we will be discussing are indicated.

## 2. Galaxies at $z \leq 1$

The association between metal–line QSO absorption systems has come a very long way in the last 15 years since the first observations in search of absorbing galaxies were attempted[30]. The connection between the absorbers and things we would call galaxies went from fairly compelling statistical and consistency arguments in the late 70's and early 80's[19], to a rather firm connection by the late 80's based on direct observations made possible using sensitive CCD cameras and spectrographs[2]. For the last

four+ years, we have been working on a comprehensive survey of Mg II absorption–selected galaxies for $z \leq 1$. Our aim has been to try to develop an understanding of field galaxies which would be parallel to the rapid developments in the faint redshift surveys[4,7,8,9,10,14,15], and to establish a starting point at redshifts where the absorbers can be compared directly with the field samples. We have been less concerned with further establishing the absorber/galaxy connection (although our large sample does accomplish this as a by–product) but instead with exploiting the unique capabilities of an absorption–selected sample to study normal galaxy evolution. Rather than repeat many of the arguments that we have already published elsewhere[23,26,27], we will concentrate instead on the bottom–line results from our now–completed survey, mostly in order to set the stage for §3 of this contribution.

The final "moderate $z$" sample consists of 58 absorbing galaxies along 48 different lines of sight, in the redshift range $0.3 \leq z \leq 1.0$. We have also observed $\sim 25$ "control" fields, which are QSO fields in which no absorption systems are found in the redshift range of interest. In addition to the deep optical and near–IR images, follow–up optical spectroscopy in both "absorber" and "control" fields has been used both to confirm the redshifts of the putative absorbers, and to discover possible "interlopers", or galaxies that are as close to the QSO line of sight as typical absorbers but which do not produce detectable absorption. A spectroscopically confirmed absorbing galaxy, or a candidate absorber whose color, impact parameter, and luminosity is fully consistent with the confirmed cases (about 70% of the sample have confirmed redshifts) has been identified in every case examined. Only a very small number of "interlopers" has been found, and these all have the property of being intrinsically very faint ($M_K \geq -22$ for $H_0 = 50$ km s$^{-1}$Mpc$^{-1}$) in absolute $K$ luminosity and also very blue in $\mathcal{R} - K$ color. There are clear correlations between the impact parameters of the identified absorbers and the strength of the observed Mg II absorption lines, as well as a correlation between identified galaxy luminosity and impact parameter that results from a size–luminosity relation (see below). The former correlation is very useful, it turns out, as a means of making a *prediction* of the impact parameter of the absorber on the basis of the absorption line properties alone. The distribution of rest–frame optical/IR colors (see Figure 2) of the absorbers generally spans the range expected for galaxies of "normal" Hubble sequence spectroscopic types ( including early–type galaxies), with the mean color remaining that of an "Sb" over the whole observed range. We also see no trend in rest–frame $B$ or $K$ luminosity or impact parameter with redshift.

We emphasize that we do not find bright "interloper" galaxies, so that the absorption phenomenon is not peculiar to some subset of galaxies, but instead appears to apply to *all* "normal" galaxies with $M_K \leq -22$ that are within a characteristic distance from the QSO line of sight

$$R(L_K) \approx 35 \left(\frac{L_K}{L_K^*}\right)^{0.2} \text{ h}^{-1} \text{ kpc.}$$

The absence of interlopers, as we have repeatedly emphasized, also implies both that the gaseous halos must be roughly *spherical* (rather than disk–like) and that the covering fraction of the diffuse gas within the characteristic radius must be ∼unity. It is very interesting to note that whether or not a given galaxy will be an absorber seems to depend more on $K$ luminosity than on rest–frame $B$ luminosity, and that the presence of an extended gaseous halo appears to be independent of the current rate of star formation as indicated by the galaxy spectrum and color. This suggests that the presence of the halo depends more importantly on galaxy *mass* than on other empirically accessible characteristics[26].

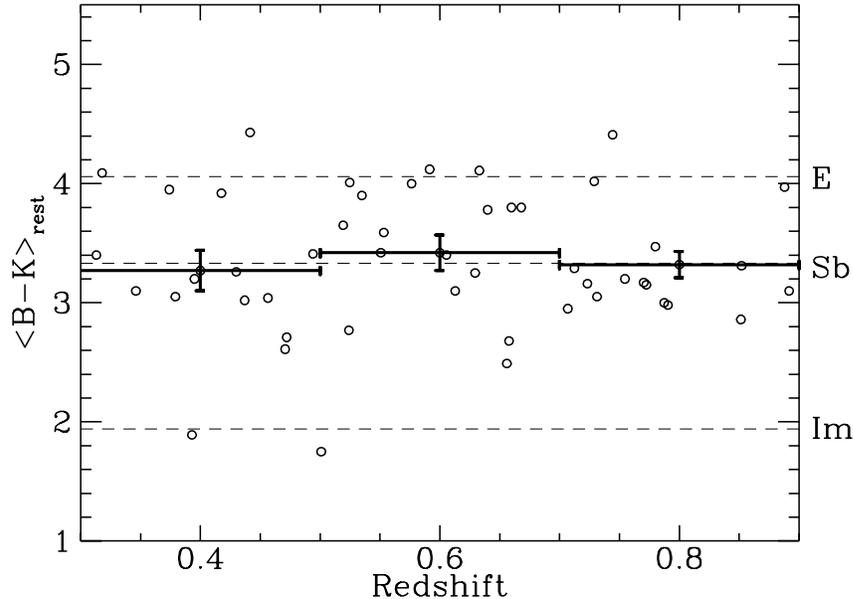

Fig. 2. Plot showing the rest–frame optical/IR color versus redshift for the moderate redshift absorbing galaxy sample (open symbols). The expected colors for unevolving galaxies or various spectroscopic types[6] are indicated with horizontal lines. The mean color in 3 redshift bins is indicated with error bars; it remains that of an "Sb" over the entire observed range.

Evolution in the galaxy population number density with redshift has been invoked in order to simultaneously reconcile the faint number counts and the observed redshift distribution in the faint field galaxy surveys[5,13,18]. It is straightforward for the absorption–selected sample to convert the observed distribution of absolute luminosities into a proper luminosity function, since the "volume" probed by the survey is only weakly dependent on luminosity ($V \propto L^{0.4}$). The normalization of the luminosity function is obtained from a combination of the observed $dN/dz$ relation from the absorption line survey and the observed distribution of "impact parameters" (from which the cross-section/luminosity relation was derived). In Figure 3 we plot both the rest–frame $B$ and rest–frame $K$ luminosity distributions (i.e., a luminosity function with arbitrary normalization). Note how well the galaxies at $\langle z \rangle = 0.65$ agree with the "present–day" $K$ band luminosity function, and that the $B$ band luminosity function of the absorbers matches the bright end of the present–day distribution reasonably well, but falls off at fainter magnitudes, with a form more Gaussian than Schechter[20]. We attempt to explain this in the following qualitative way: if one looks at the distribution of a local "field" sample broken down by morphological type[3], the $B$ luminosity function of galaxy types earlier than Sd actually bears very close resemblance to that which we observe for the absorbing galaxies. The faint end of the local luminosity function is dominated by galaxies of late morphological type (aside from dwarf ellipticals) which tend to have very blue colors. We see a very clear trend in our absorber sample of $M_K$ and $(B - K)_0$, in the sense that fainter galaxies have bluer colors, whereas we do not see any such trend in $M_B$ versus $(B - K)_0$. As discussed above, it is the $K$ luminosity that seems to be the major factor governing whether or not a galaxy is an absorber. Thus there is a population of galaxies which are intrinsically faint at $K$ but with substantial $B$ luminosities (i.e., they are very blue) that we are "missing" in the absorber sample. This is consistent with the observed nature of the "interloper" galaxies found in our absorber and control fields. All of this implies that there may be, in addition to an apparent dependence of gaseous halos on galaxy *mass*, a morphological selection effect as well. Absorption selection picks out relatively massive galaxies, of "normal" Hubble type, and selects against small, blue galaxies with less substantial mass and (presumably) more "irregular" morphology.

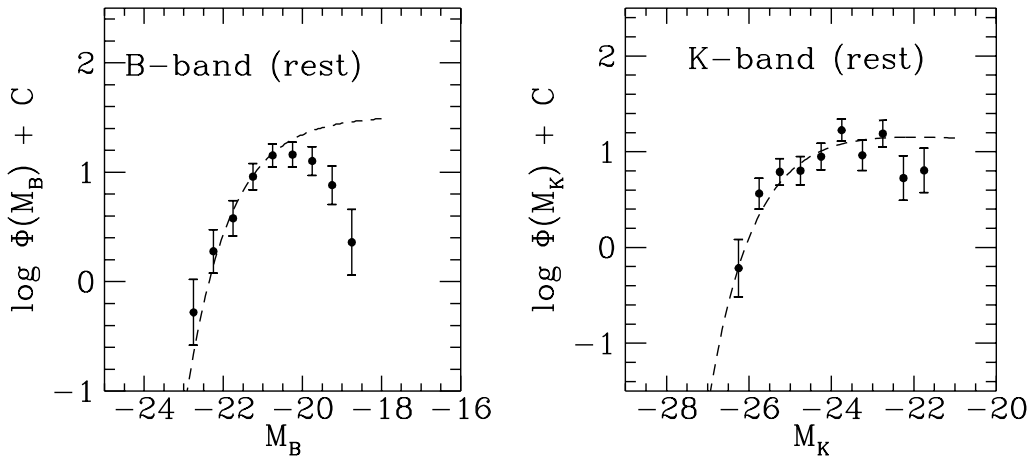

Fig. 3. The rest–frame $B$ and $K$ luminosity functions for the moderate redshift absorbing galaxy samples. Recent determinations of the "local" $B$[16] and $K$[17] luminosity functions for field galaxies are shown with the dashed curves. Note the Gaussian appearance of the $B$ luminosity function while the $K$ luminosity function for the same galaxies is well-fitted by a Schechter function. The normalization has been left as a free parameter (see text).

The implications of the actual space density inferred for the absorbing galaxy population are very interesting. We obtain a normalization as described above,

$$\Phi^* = 3.0 \pm 0.7 \times 10^{-2} \quad \text{Mpc}^{-3}$$

(for $H_0 = 100$ km s$^{-1}$Mpc$-1$ and $q_0 = 0.05$; adopting $q_0 = 0.5$ makes the required space density about 15% higher), which is about a factor of 2 higher than the most recent determination for a sample of local field galaxies[16]. The curious thing is that we see no evidence for an increase in the space density of the galaxies over the redshift range of our survey; since $dN/dz$ is consistent with no change in total cross-section (Figure 1), and there is no trend of impact parameter with redshift (see discussion in §3), this forces us to the conclusion that, if indeed the difference in normalization is significant, then essentially all of the evolution in space density must have occurred between $z \sim 0.3$ and $z \sim 0$ (this becomes even more curious in light of our discussion in §3).

The bottom line: the absorbing galaxy population, which appears to be the population of galaxies having "normal" Hubble type and substantial $K$ luminosity, shows no evidence for significant color, luminosity, or space density evolution since $z \sim 1$. With the exception of a persistent factor of $\sim 2$ normalization "problem", these galaxies appear in every way to be consistent with their present–day counterparts. This can be contrasted with the remarkable evolution seen in the B–selected surveys (see M. Colless's contribution to these proceedings). We can only conclude that there are two distinct galaxy populations, one of which must be evolving very rapidly while the other remains stable. The evolution of the "faint blue galaxies" apparently goes largely unnoticed by the more massive galaxies traced by our survey. These conclusions appear to be in broad agreement with the initial findings of the HST "Medium Deep Survey."[12]

## 3. Galaxies at $z > 1$

From Figure 1 it is easy to see that maintaining a consistent selection criterion to push the survey out to higher redshift is straightforward; however, the galaxy identification suffers to some extent from the same limitations as the field galaxy redshift surveys. We can, however, use the results described above for galaxies at $z \leq 1$ to make "no evolution" predictions about what the galaxies should look like at higher redshifts, in terms of their apparent magnitudes, colors, impact parameters, etc. It is possible to make very strong statements about the evolution of the absorption–selected galaxies *without performing follow–up spectroscopy.* The key to success in performing a statistical identification of the absorbing galaxies is the set of concrete predictions which result from the work at $z < 1$ *and* comprehensive knowledge of the absorption line systems *at all redshifts* along each QSO sightline. Our target redshift range has been $1.0 \leq z \leq 1.6$, in order to produce a sample having a mean redshift of $\langle z \rangle = 1.3$. The basic data consist of deep optical (to $\mathcal{R} = 26$) and near-IR (to $K \sim 21.5$) images of a carefully chosen set of fields, selected using identical criteria to our moderate redshift survey, and for which the absorption line spectroscopy is exceptionally complete.

In practice, we can choose lines of sight which have one, two, or even three absorbers in the redshift range of interest, as well as relatively rare "control" fields in which no Mg II absorption systems are present from $z \sim 0.2$ to the QSO emission redshift (typically $z \sim 2$). This method allows us to predict *a priori* how many candidates we should find in a given field, and allows us to assess how important the contamination by foreground, blue "interloper" galaxies might be (note that the optical/IR colors of such galaxies are *far* bluer than the expectations for the high redshift absorbing galaxies–see Figure 4). We know from our moderate redshift survey that, in the absence of evolution, we expect, e.g., typical apparent $\mathcal{R}$ magnitudes of 24–26, impact parameters in the range $0'' \leq \theta \leq 7''$ (with the most probable value being $4\rlap{.}''5$), and typical $\mathcal{R} - K$ colors of $\sim 4 - 5$. Any departure from the "no evolution" expectations indicates significant evolution of the galaxy population.

At $z = 1.6$ our optical passband samples rest–frame 2670Å, so that the $K$-band observations are absolutely essential in order to be complete irrespective of current star formation rates, and they also allow a much more robust comparison of absolute rest–frame luminosities of our high redshift and moderate redshift galaxy samples because the differential k–corrections are small and largely independent of galaxy type.[1,27]

To date we have complete data for 12 absorber fields (containing 19 absorbing galaxies) and 3 control fields; this constitutes about 30% of the full sample. Thus far, we have been able to identify candidate absorbing galaxies within the expected angular separation of the QSO line of sight in every case (although, as we shall discuss, some galaxies are thus far detected *only* in the near-IR). In cases where there is more than one absorbing galaxy identified, provisional redshift assignments are accomplished by using the empirical relation between absorption line strength and impact parameter from our moderate redshift survey[27]; thus the scatter introduced by uncertainties in redshifts for individual galaxies is significantly reduced as compared to random guessing.

In Figure 4 we have plotted $\mathcal{R} - K$ colors versus redshift for both the moderate and high redshift samples. Note first of all that the colors of the identified high redshift absorbers, relative to the unevolved model spectral energy distributions[6], are very much in line with what was predicted on the basis of the moderate redshift sample. The mean color, as at lower redshift, is consistent with an *unevolved* mid–type spiral galaxy. There is also roughly the same *range* of color represented in the high redshift sample as at moderate redshift, although we do see a potential "blueing" in the red envelope relative to the elliptical model SED; whether or not this effect is significant will become more evident when we have completed our full sample.

In connection with the colors and magnitudes of the $z > 1$ objects, it is particularly interesting to place the galaxies in the $\mathcal{R} - K$ versus $K$ color–magnitude diagram, for direct comparison with the various field galaxy redshift surveys[10,11,21] (Figure 5). From the diagram it is clear that if substantial numbers of $z > 1$ galaxies are to be observed in field galaxy redshift surveys, then the searches should be directed at objects having $K > 19$ and $\mathcal{R} - K \gtrsim 4$. The problem, once again, is that the galaxies are extremely faint optically; note that it is exactly this part of the color–magnitude diagram that is spectroscopically incomplete in the deepest $K$–selected redshift survey.[10,21]

What about the luminosities and the space density of the absorbing galaxy popu-lation beyond $z = 1$? In Table 1 we list some of the relevant statistics for the moderate and the high–redshift samples. The absolute $K$ luminosity (converted to rest–frame) allows a direct comparison between the moderate and high redshift samples. Both the mean value and the distribution of absolute $K$ luminosities are different at the $\sim 2\sigma$ level if $q_0 = 0.05$, in the sense that galaxies are slightly brighter at $K$ in the high redshift subsample. However, if $q_0 = 0.5$, then both the luminosity distribution and the mean luminosity are statistically equivalent in the two subsamples. Thus, for high $q_0$, there has been no luminosity evolution whatsoever over the interval $1.6 \geq z \geq 0.3$! (This brings up the important point that may be overlooked in analyses of the luminosity evolution of high redshift galaxies: for $z > 1$ it is very difficult to discriminate between luminosity evolution and one's choice of cosmology.) In fact, population synthesis models predict at least some luminosity evolution over such a large span of time, just from passive evolution of the stellar populations comprising the $K$ band light[6].

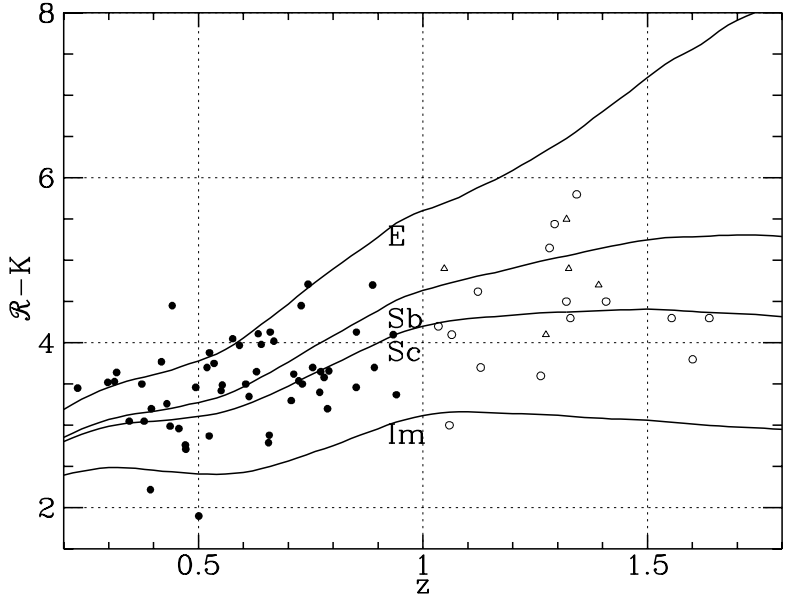

Fig. 4. Plot showing the optical/IR color versus redshift for the absorbing galaxy samples (solid and open symbols). Open triangles indicate galaxies with only lower limits on $\mathcal{R} - K$. The curves are the expected tracks of *unevolved* model galaxy spectra from Bruzual and Charlot (1993). Note that the mean color remains that of a mid–type spiral to the highest redshift observed; also note the general absence of very blue galaxies from both subsamples.

As for the space density of the galaxies: we have already seen that $dN/dz$ for the absorbers is consistent with the product of the galaxy space density and mean cross-section remaining constant over the redshift range $0.3 \leq z \leq 2.2$. It is therefore possible to make a direct comparison of galaxy space densities at various redshifts by looking at the mean cross-section of the absorber population to see if it is changing significantly. An empirical measure of this is the mean value of the square of the observed impact parameter, as given in Table 1. It is clear that there is no significant difference in these values between the two redshift subsamples; if we include all of the uncertainties involved in the value of $dN/dz$ and $\langle D^2 \rangle$ and a range of possible values of $q_0$, then the space density of the absorbing galaxy population cannot have changed by more than $\sim 30\%$ over the entire redshift range spanned by our survey, $1.6 \geq z \geq 0.3$. We therefore find it somewhat "suspicious" that the normalization, observed to be essentially constant over a very large range of redshift, can have changed by a factor of $\sim 2$ since $z \sim 0.3$!

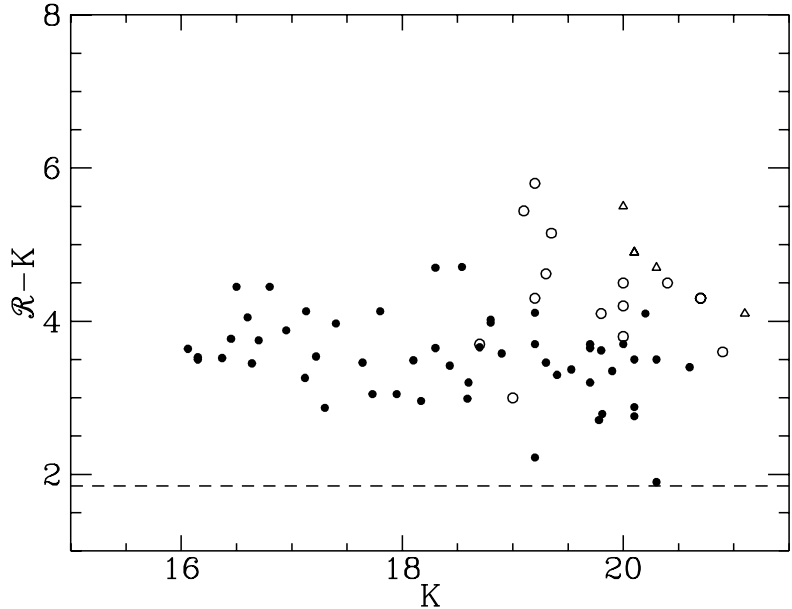

Fig. 5. Color magnitude for the absorbing galaxy sample for $z < 1$ (solid symbols) and $z > 1$ (open symbols). The open triangles indicate galaxies for which only lower limits exist on $\mathcal{R} - K$. Note that the bulk of the high redshift galaxies are found at $K > 19$ and $\mathcal{R} - K > 4$.

## 4. Discussion and Future Progress

It is clear that at $z \leq 1$, gas cross-section selects the population of apparently normal galaxies which have $M_K \lesssim -22$, or $L_K \gtrsim 0.07 L_K^*$; generally absent from the sample are galaxies which are both intrinsically faint in the near-IR, which happen to be also very blue in their optical/IR colors. The implication is that these latter galaxies, which appear in large numbers particularly in surveys selected in the observed $B$ band[9], do not possess extended envelopes of diffuse gas. The sample of absorbers appears to include both early and late type galaxies; the fact that the nature of the halos is uncorrelated to first order with the current rate of star formation is (we note in passing) strong evidence *against* the idea that the diffuse halos are produced by gas expelled from the galaxy by supernovae in a giant Galactic fountain. The fact that all galaxies of a particular $K$–band luminosity (suggesting a particular stellar mass) possess these extended halos of diffuse gas, and that this population of galaxies appears to have been stable over a long period of time, suggests both an early epoch of formation for relatively massive systems, *and* a prolonged period of *accretion* of gas from the galaxies' surroundings. The fact that there are at most small changes in the *rest frame* optical/IR color of the population as a whole suggests that the star formation rate of the overall population remains essentially constant.

Table 1. Absorbing Galaxy Luminosity and Size Parameters

| Statistic | $q_0 = 0.05$ | $q_0 = 0.5$ |
|---|---|---|
| | $\langle z \rangle = 0.65$ | |
| $\langle M_B \rangle$ | $-20.64 \pm 0.12$ | $-20.35 \pm 0.12$ |
| $\langle M_K \rangle$ | $-24.03 \pm 0.16$ | $-23.74 \pm 0.16$ |
| $\langle D^2 \rangle^{1/2}$ | $25.2 \pm 1.2$ h$^{-1}$ kpc | $22.3 \pm 1.1$ h$^{-1}$ kpc |
| | $\langle z \rangle = 1.30$ | |
| $\langle M_K \rangle$ | $-24.45 \pm 0.14$ | $-23.87 \pm 0.14$ |
| $\langle D^2 \rangle^{1/2}$ | $25.4 \pm 1.6$ h$^{-1}$ kpc | $20.7 \pm 1.3$ h$^{-1}$ kpc |

It is (to us) fascinating that, even to $z \sim 1.6$, the properties of the galaxies selected using gas cross-section are not radically different in color, luminosity, or space density from their counterparts at much lower redshift (and, perhaps, even their counterparts of the present epoch). This finding goes somewhat against a tendency during the last 10 years or so to think of galaxies as we know them to be a very recent phenomenon, having changed radically in number and star formation rate in the relatively recent past. It has turned out to be very important to specify, when making broad statements about galaxy evolution, *which* galaxy population you are talking about. This brings up the very important question which it appears we now need to answer: how are the relatively quiescent, massive galaxies, apparently unchanged in any radical way since $z \sim 1.6$, related to the population of what has become known as the "faint blue galaxies", which apparently evolve at a fantastic rate? One possibility that would simultaneously explain many of the observations in a qualitative way is that the "small" galaxies disappear over time by being accreted onto the larger galaxies, and that the gaseous halos of the massive galaxies have their origin (at least in part) in tidally disrupted gas from the faint galaxies (timescale arguments require that the infalling "halo" gas be continually replenished over a very long period of time)[22,28,29] . This hypothesis would predict that the many of the "faint blue galaxies" should be gravitationally associated with brighter galaxies that are members of the absorbing galaxy population, so that it can be tested by performing deep redshift surveys (using

multi–object spectroscopy) in the fields of the QSO absorbers (which we plan to undertake in the near future). In any case, understanding the environment of the absorbers should go a long way toward explaining the origin of the extended halos.

Another question has come to the forefront as well: is there really a normalization problem for the "normal" galaxy population? This factor of $\sim 2$ discrepancy found by many surveys at moderate redshift as compared to zero redshift is somewhat disturbing, considering that the normalization does not seem to change at nearly the same "rate" when the surveys are extended to still higher redshift. It would seem to be very important to understand whether there are possible systematic effects in the "local" surveys that might result in a luminosity function normalization that is too low. The revolutionary wide–field surveys discussed by others at this meeting will almost certainly settle the issue.

Finally, there is the question of whether we can trace the population of relatively massive galaxies back to an epoch when they actually *do* look substantially different. This can certainly be attempted using the kinds of techniques we have discussed above, applying the same selection criteria to follow galaxies beyond $z \sim 2$. Observations in the near–IR are absolutely crucial for making epoch–to–epoch comparisons. If there is an epoch at which the normal galaxies, e.g., have not yet assembled from a group of "sub-units", one should be able to find it using deep optical and near–IR imaging alone, provided that adequate care is taken to observe control fields and understand the extent of "contamination" by interlopers. Also, at $z > 3$, other complementary methods can be used for identifying a population of normal galaxies, and in fact it is possible to use essentially identical selection criteria as at lower redshift to target the search[24,25].

The truly exciting prospect for obtaining a deeper physical understanding of the nature of these extremely distant galaxies is that *every one of them* can be studied in more detail by obtaining high resolution spectra of the background QSOs using the new generation of 8-10m telescopes equipped with echelle spectrographs. The complementary information which would come from both direct identification of the galaxies and the accompanying high resolution spectroscopy, all as a function of redshift and accumulated using well–defined samples, will allow us a much more intimate view of the process of galaxy evolution (and, perhaps, galaxy formation) than is otherwise accessible.

C.C.S. would like to thank the Royal Greenwich Observatory and the Institute of Astronomy for their generosity in supporting a visit that included the Herstmonceaux meeting. We gratefully acknowledge the patience and generosity of the time assignment committees of Lick Observatory and the Kitt Peak National Observatory, without which the work reported above would not have been possible.